\documentclass[12pt, a4paper]{article}
\usepackage[utf8]{inputenc}
\usepackage[T1]{fontenc}
\usepackage[babel=true]{microtype}
\usepackage{amsmath}
\usepackage{amsfonts}
\usepackage{amssymb}
\usepackage{graphicx}
\usepackage{geometry}
\usepackage{booktabs}
\usepackage{array}
\usepackage{multirow}
\usepackage{caption}
\usepackage{setspace}
\usepackage{longtable}
\usepackage[colorlinks=true, linkcolor=blue, citecolor=red, urlcolor=cyan]{hyperref}
\usepackage{doi}

\title{Validation of Product Nuclide Activity Calculations in IAEA Charged-Particle Cross Section Database for Beam Monitor Reactions}

\author{
    Mustafa Rabuş\textsuperscript{1*} 
    \and 
    İskender Atilla Reyhancan\textsuperscript{2}
}
\date{}
\begin{document}
\maketitle
\noindent
\textsuperscript{1}Istanbul Technical University, Energy Institute, İstanbul, Türkiye \\
\textsuperscript{2}Istanbul Technical University, Energy Institute, Nuclear Researches Division, İstanbul, Türkiye

\vspace{0.3cm}
\noindent
*\textbf{Corresponding author:} \texttt{rabus18@itu.edu.tr}
\vspace{0.5cm}

\renewcommand{\baselinestretch}{1.5}
\large
\textit{This manuscript is currently under review in Journal of Radioanalytical and Nuclear Chemistry.}
\normalsize

\begin{abstract}
For the 34 monitor reactions included in the IAEA Beam Monitor Reactions (BMR) 2017 dataset and the 22 reactions listed in the IAEA BMR 2007 dataset, product radionuclide activities were calculated using the IMRA computational framework developed in this work. The aim of this work is to independently validate radionuclide activity calculations for charged-particle monitor reactions by comparing results obtained with the IMRA computational approach against the reference activity data provided in the IAEA-BMR 2007 and 2017 datasets. The calculated activity values were then compared with the corresponding pre-calculated activity data provided in the IAEA BMR datasets.

This comparison demonstrates overall consistency between the present calculations and the IAEA reference data across the evaluated monitor reactions. However, for a limited subset of monitor reactions induced by doubly charged particles ($\alpha$ and $^3$He), notable differences were observed when the IAEA BMR 2017 activity values were used. These observations emerged during the independent validation of the present computational methodology implemented in the IMRA code and are reported as part of a reliability assessment of the activity calculation procedure.

\end{abstract}

\section{Introduction}

Radioisotope production processes are fundamentally based on nuclear interaction cross-section data. To optimize production conditions, the target material, incident projectile, beam energy, beam intensity, and irradiation time must be carefully selected. In this context, it is important to distinguish between the number of produced radionuclide nuclei and their resulting activity, which depends on both the irradiation history and radioactive decay properties \cite{qaim2002charged,IAEA2001}.

For beam monitor reactions, radionuclide activity does not represent an end-use quantity as in medical or industrial applications. Instead, calculated activity values serve as a reference quantity for the derivation of experimental cross-sections and for beam fluence calibration, as established in standard monitor reaction evaluations \cite{takacs2007validation,hermanne2018reference,Tarkanyi2025}. Within this framework, the accuracy of calculated activity values is therefore of methodological importance and provides a suitable basis for testing the reliability and internal consistency of activity calculation procedures.

In this work, radionuclide activity values were calculated for 34 monitor reactions included in the 2017 version \cite{IAEA_BMR_2019} and for 22 monitor reactions listed in the 2007 version \cite{IAEA_BMR_2007} of the International Atomic Energy Agency (IAEA) Beam Monitor Reactions (BMR) dataset, using independently developed computational methods implemented in the Ion-Matter Range and Activity (IMRA) code.

During the independent validation of this computational approach, notable differences were observed for a limited subset of monitor reactions induced by doubly charged particles ($\alpha$ and $^3$He) when activity values derived from the IAEA BMR 2017 dataset were used. To further assess the robustness of these observations, the same calculations were applied to the corresponding reactions in the IAEA BMR 2007 dataset, for which such differences were not observed within the same computational framework.

\section{Computational Methods}
\subsection{Mathematical Procedure}
As a charged particle beam propagates through matter, nuclear reactions occur with a probability governed by the energy-dependent nuclear cross section. In the present approach, the beam energy is discretized into fixed energy steps of 0.1~MeV. Owing to the stopping power described by the Bethe equation, each energy decrement corresponds to a spatial displacement step, \(dx\), whose length varies with energy.

At each energy step, the particle beam undergoes a displacement at a different penetration depth. Although the atomic number density of the target material remains constant, the particle beam encounters a different number of target atoms within each displacement step due to the varying length of the corresponding \(dx\) segment. For this reason, within the IMRA computational approach, the Bethe equation is employed in conjunction with the activity equations. The activity calculations are further detailed in the subsequent sections.

\subsection{Bethe Formula Application for Beam Step Depth Calculation}
The theoretical basis for the interaction of fast charged particles with matter was established in the pioneering work of Bethe \cite{bethe1930theorie}. The calculation of stopping power, arising from energy loss through ionization, excitation, and electromagnetic interactions, can be performed using the Bethe formula. The Bethe formula provides the average energy lost (dE) per unit path length (dx) in the material \cite[pp. 100--103]{tsoulfanidis2015measurement}. The stopping power expression for protons, deuterons, tritons, and alpha particles is given in Equation 1. The CSDA (Continuous Slowing-Down Approximation) range can then be evaluated using Equation 2 \cite[p. 30]{IAEA_Cyclotron_Targets}.

\begin{equation}
-\frac{dE}{dx}=\frac{4\pi N_{A}r_{e}^{2}m_{e}c^{2}}{\beta^{2}}\cdot\frac{Z}{A} \cdot z^{2}\left[\ln\left(\frac{2m_{e}c^{2}\beta^{2}\gamma^{2}}{I}\right)-\beta ^{2}\right]
\end{equation}

\begin{equation}
R_{\text{CSDA}}=\int_{0}^{E_{0}}\left(\frac{dE}{dx}\right)^{-1}dE
\end{equation}

In Equation (1), the symbols and physical constants have their standard meaning within the Bethe stopping power formalism and are defined as follows:

\begin{itemize}
    \item $-\dfrac{dE}{dx}$ is the stopping power, representing the mean kinetic energy loss of the incident charged particle per unit path length in the target material.
    \item $z$ is the charge number of the incident particle in units of the elementary charge.
    \item $Z$ and $A$ denote the atomic number and atomic mass of the target material, respectively.
    \item $N_A$ is Avogadro’s number.
    \item $r_e$ is the classical electron radius.
    \item $m_e$ is the rest mass of the electron.
    \item $c$ is the speed of light in vacuum.
    \item $\beta = v/c$ is the reduced velocity of the incident particle, where $v$ is its instantaneous velocity.
    \item $\gamma = (1-\beta^2)^{-1/2}$ is the corresponding Lorentz factor.
    \item $I$ is the mean excitation (ionization) potential of the target material.
\end{itemize}

In Eq.~(2), $E_0$ represents the initial kinetic energy of the charged particle beam at the entrance of the target. In the general formulation of the Bethe stopping power formalism, the lower integration limit corresponds to zero kinetic energy, i.e.\ complete stopping of charged particles in the material. However, in the present work the calculations are restricted to the energy interval covered by the IAEA-BMR reference data. Consequently, the integration is performed from $E_0$ down to the lowest energy value available in the database, rather than to zero kinetic energy.

In Eq.(3) discrete summation approximates the integral in Eq (2);

\begin{equation}
R_{\mathrm{CSDA}} = \sum_{E_{\mathrm{MIN}}}^{E_{\mathrm{MAX}}} 
\frac{\Delta E}{S(E)}
\end{equation}

\noindent
where:

\begin{itemize}
    \item $R_{\mathrm{CSDA}}$ : CSDA (Continuous Slowing Down Approximation) range, the total path length of a particle assuming continuous energy loss, in meters (m).  
    \item $S(E)$ : stopping power at energy $E$ (energy loss per unit path length), in units of MeV/m.  
    \item $\Delta E$ : discrete energy step used in the summation, in MeV.  
    \item $E_{\mathrm{MIN}}, E_{\mathrm{MAX}}$ : minimum and maximum particle energies considered, in MeV.  
\end{itemize}

Here, $dE$ and $\Delta E$ denote the energy loss per numerical step. Following the recommendations given in technical reports from the IAEA, the energy decrease per step was set to $\Delta E = 0.1~\text{MeV} $ \cite[p.~30]{IAEA_Cyclotron_Targets}.
The displacement steps and the CSDA range (Equation 3) were computed numerically using the IMRA code developed within the scope of this work.

A comparison between the CSDA ranges calculated in this work and the projected ranges derived from SRIM \cite{srim_web} showed discrepancies of less than 4 percent for all reaction channels. The percentage deviation of the range values calculated in this study from the SRIM code is shown in Table\ref{tab:range_comparison}. The input and output text files generated from the SRIM calculations for the IAEA-BMR datasets are accessible via Zenodo \cite{Rabus2025_SRIM_BMR}.
The differences observed between the present calculations and SRIM results arise primarily from differences in the underlying calculation approaches. SRIM evaluates ion transport using a statistical treatment of scattering processes and reports projected ranges along the beam direction, whereas the present approach is based on the Bethe stopping power and evaluates the CSDA path length along the beam trajectory. Furthermore, a systematic tendency is suggested by the range comparisons: as the atomic mass of the target material and the mass of the incident ions increase, the projected ranges obtained from SRIM are observed to be slightly larger than the CSDA ranges calculated with the IMRA approach. Conversely, for interactions involving lighter target nuclei and lower-mass projectile ions, the CSDA ranges derived from IMRA appear to be marginally larger than the corresponding SRIM projected ranges. In addition, small numerical differences may also result from variations in the input parameters and numerical precision used in the respective implementations. These methodological differences can reasonably account for the small deviations observed between SRIM and IMRA results in Table\ref{tab:range_comparison}.

While CSDA range and beam penetration data for energetic charged particles can be obtained with high precision, these quantities are not the final objective of the present study. In the IMRA approach, the Bethe stopping power is used to determine successive displacement steps, \(\Delta x\), corresponding to fixed energy losses (\(\Delta E = 0.1~\mathrm{MeV}\)) along the particle trajectory. These stepwise displacement lengths are subsequently employed in the activity calculations, where radionuclide activity values are evaluated incrementally along the particle path.

\subsection{Nuclear Activation Dependence on Beam Flux Variation and Target Nuclide Variation per Step}

The particle flux ($\phi$) was calculated at each displacement step $\Delta X_i$ using the cross-section value $\sigma_i$ corresponding to each energy $E_i$ according to Eq.~6. The depletion of product radionuclides due to secondary interactions can be neglected, as the probability of such interactions is extremely small. Following this approximation, the cross-section of the product radionuclides is considered to be zero. The stepwise form of Eq.~(4) was derived specifically for the IMRA implementation and is documented in detail in Ref.~\cite{rabus2020pozitron}, while the underlying
activation formalism follows the standard treatment given in Ref.~\cite{lamarsh2001introduction}.

\begin{equation}
A(t)
=
\lambda
\left[
\frac{\phi \sigma N_{0}}{\lambda - \phi \sigma}
\left(
e^{-\phi \sigma t}
-
e^{-\lambda t}
\right)
\right]
\end{equation}

where
\begin{itemize}
    \item \(A(t)\) : activity of the produced radionuclides at time \(t\) (Bq),
    \item \(\lambda\) : decay constant of the product radionuclide (s\({}^{-1}\)),
    \item \(\phi\) : particle flux (m$^{-2}$\,s$^{-1}$),
    \item \(\sigma\) : reaction cross section (m\({}^{2}\)),
    \item \(N_{0}\) : initial target nuclide quantity,
    \item \(t\) : irradiation time (s).
\end{itemize}

The discrete computational form of the activation equation, which accounts for energy-dependent variations, is:

\begin{equation}
A(E,t)
=
\sum_{i=0}^{E/\Delta E}
\left[
\frac{
\lambda \, \phi_i \, \sigma(E_i)\, N_i
}{
\lambda - \phi_i \sigma(E_i)
}
\left(
e^{-\phi_i \sigma(E_i)\, t}
-
e^{-\lambda t}
\right)
\right]
\end{equation}

The particle flux at a position $x$ along the particle path, accounting for attenuation due to cumulative interactions, is given by:

\begin{equation}
\phi(x) = \phi_0 \, 
\exp\Bigg[ 
- \sum_{i=1}^{E/\Delta E} \sigma(x_i) \, \Delta x_i \, N_v 
\Bigg],
\end{equation}
where $\phi_0$ is the initial flux at the entrance of the target, $\sigma(x_i)$ is the cross-section at the $i$-th spatial step, $\Delta x_i$ is the step length, and $N_v$ is the number density of target nuclei. 

The spatial step corresponding to the energy interval $\Delta E_i$ is obtained from the stopping power as:
\begin{equation}
\Delta x_i = \frac{\Delta E_i}{S(E_i)} .
\end{equation}

These \(\Delta x_{i}\) steps define the interaction volumes required for activity calculations, as they determine the number of target atoms (\(N_{i}\)) contributing to the reaction at each energy value. Since the displacement length \(\Delta x_{i}\) varies with energy, this variability is explicitly taken into account in the calculation. The atomic number density of the target material is given by:
\begin{equation}
N_v = \frac{d\,N_A}{A},
\end{equation}
where $d$ is the mass density and $A$ is the molar mass of the target.

The target nuclei within the $i$-th energy interval segment are then given by:

\begin{equation}
N_i = N_v \, \Delta x_i \cdot (1~\mathrm{m^2}),
\end{equation}

Here, the reference area is a normalization convention and not a physical geometric area. This approach is consistent with the fact that the nuclear reaction cross-section $\sigma$ has units of area (m²) and represents the effective interaction probability of an incident particle with a single target nucleus.

In this context, the energy domain is discretized into finite intervals of width $\Delta E$. The index $i$ labels the corresponding energy step, with $E_i$ denoting the representative energy of the interval. The particle flux $\phi_i$ is evaluated at the entrance of the spatial step $\Delta x_i$ corresponding to the $i$-th energy interval. The quantity $N_i$ represents the number of target nuclei contained within the spatial displacement step $\Delta x_i$.

Product radionuclide yields and activity values obtained without neglecting this stepwise variation provide a more realistic representation of the activation process. A schematic illustration of the stepwise displacement segments ($\Delta x_i$) defining the interaction volumes along the beam path is shown in Fig.1.

\begin{figure}[htbp]
\centering
\includegraphics[width=1\textwidth]{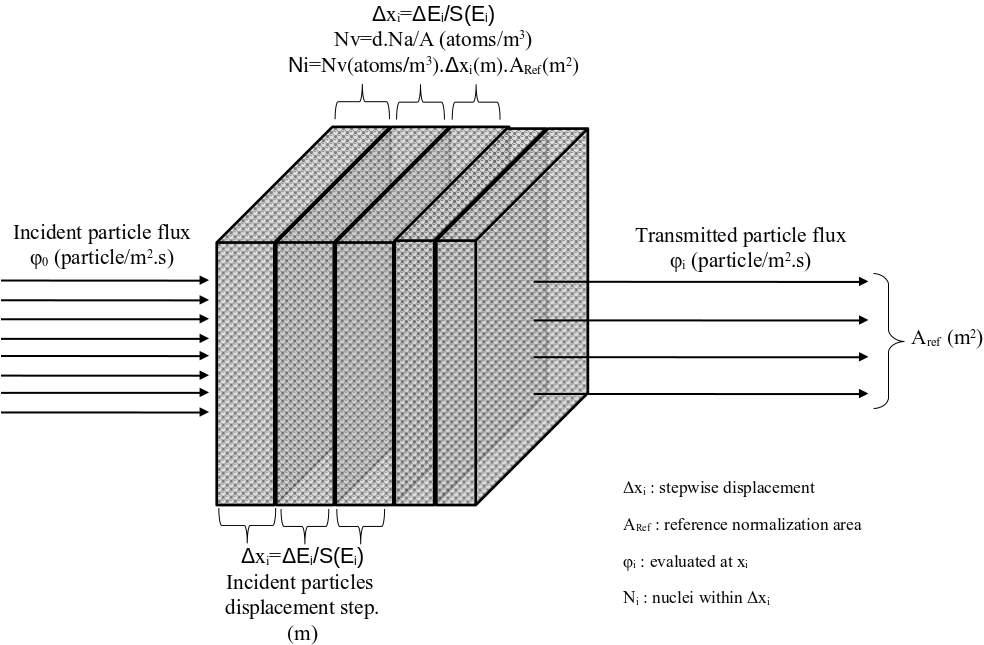}
\caption{Schematic illustration of the stepwise displacement segments ($\Delta x_i$) defining the effective interaction volumes along the beam path. Each segment corresponds to a finite energy interval $\Delta E$, where the particle flux $\phi_i$ is evaluated at the entrance of the spatial step and $N_i$ denotes the number of target nuclei contained within $\Delta x_i$. The reference area shown in the schematic serves solely as a normalization convention and does not represent a physical geometric area.}
\label{fig:stepwise_geometry}
\end{figure}

\section{Simulation Code}

The calculations presented in this study were performed using the IMRA (Ion--Matter Range and Activity) code \cite{Rabus2025_IMRA_BMR}, a deterministic simulation software developed in the \texttt{C++} programming language. IMRA is based on original algorithms implementing the mathematical procedures described in the previous sections and is designed to model the interaction of energetic charged particles with matter, their penetration behavior within the target, and the resulting activity of product radionuclides.

The computational algorithm implemented in IMRA can be summarized as follows:

\begin{itemize}
    \item For each discrete energy loss step, $\Delta E$, the corresponding displacement step $\Delta x_i$ is calculated using Eq.~(7), while the CSDA range is evaluated using Eq.~(3).
    
    \item The number of target nuclei ($N_i$) within the $i$-th energy interval and the corresponding displacement step $\Delta x_i$ is calculated using Eq.~(9).
    
    \item The quantity of product radionuclides formed at each displacement step is determined by accounting for the energy-dependent nuclear cross section and the local target nuclide density, as expressed by the term $\phi_i \sigma(E_i) N_i$ in Eq.~(5).
    
    \item Radionuclide activity values are calculated both locally (stepwise) and cumulatively over the full particle range through the stepwise numerical integration of Eq.~(5).
\end{itemize}

The IMRA source code, together with the input and output files used in the calculations, is publicly available and documented in \cite{Rabus2025_IMRA_BMR}.

\section{Outputs}

The 34 reactions included in the IAEA BMR 2017 dataset also encompass the 22 reactions contained in the 2007 dataset, apart from some differences in the covered energy ranges. For the reactions in the IAEA BMR 2017 and 2007 datasets, radionuclide activity values were calculated under identical irradiation conditions (1~hour, 1~$\mu$A). Table~2 presents a comparison between the activity values calculated in this study and those derived from the energy-dependent cross-section data reported in the IAEA BMR 2017 and 2007 datasets.

As shown in Table~2, for alpha- and $^3$He-induced monitor reactions, the activity values reported in the IAEA--BMR 2017 dataset are approximately a factor of two higher than those calculated in the present study. In contrast, for the same reactions, the activity values reported in the IAEA--BMR 2007 dataset are generally consistent with the present calculations, with differences remaining below 5\%. This marked reduction in the observed differences may be regarded as an indication of the internal consistency of the activity calculations performed in this work.

In addition, as shown in Table~1, the differences between the SRIM projected ranges and the CSDA ranges calculated using the IMRA tool developed in this study remain below 4\%.

After excluding anomalous reactions, the mean absolute percentage error (MAPE) and the population standard deviation of the percentage differences (STDEV.P) were found to be below 5\% for all evaluated cases. This confirms that both the range and activity calculations fall within the expected uncertainty limits. Pearson correlation coefficients exceeding 0.998 indicate an almost perfect linear relationship between the IMRA model predictions and the database values. Furthermore, for range calculations, the comparison between IMRA and SRIM results yields Pearson coefficients of approximately 0.99 and MAPE values around 1.0 for both the 2007 and 2017 datasets, indicating strong consistency.

As can also be seen from Table~2, when compared with the 2007 data, the differences generally remain below 3\%, with a maximum deviation of approximately 14\%. This deviation may be related to the high sensitivity of the activity calculation equations used in this study to the cross-section inputs.

\begin{table}[htbp]
\centering
\scriptsize
\setstretch{1}
\caption{Comparison of Projectile Particle Ranges from IAEA BMR 2007 and 2017 Databases}
\label{tab:range_comparison}
\begin{tabular}{p{2.3cm}ccc p{2.3cm}ccc}
\toprule
\multicolumn{4}{c}{\textbf{IAEA BMR 2017 DATABASE}} & \multicolumn{4}{c}{\textbf{IAEA BMR 2007 DATABASE}} \\
\cmidrule(r){1-4} \cmidrule(l){5-8}
\multirow{2}{*}{\textbf{Reaction}} & \multicolumn{2}{c}{\textbf{Range (m)}} & \multirow{2}{*}{\textbf{\% Dif.}} & \multirow{2}{*}{\textbf{Reaction}} & \multicolumn{2}{c}{\textbf{Range (m)}} & \multirow{2}{*}{\textbf{\% Dif.}} \\
\cmidrule(r){2-3} \cmidrule(l){6-7}
 & \textbf{SRIM} & \textbf{This Work} & & & \textbf{SRIM} & \textbf{This Work} & \\
\midrule
natAl(a,x)\textsuperscript{22}Na & 0.00605847 & 0.00606701 & 0.141 & natAl(a,x)\textsuperscript{22}Na & 0.00276063 & 0.00276 & -0.023 \\
150 → 31 & & & & 100 → 30 & & & \\
natAl(a,x)\textsuperscript{24}Na & 0.00766847 & 0.00767874 & 0.134 & natAl(a,x)\textsuperscript{24}Na & 0.00276063 & 0.00275917 & -0.053 \\
169.9 → 31 & & & & 100 → 30 & & & \\
natCu(a,x)\textsuperscript{65}Zn & 0.00031299 & 0.00030237 & -3.395 & natCu(a,x)\textsuperscript{65}Zn & 0.00030793 & 0.00029759 & -3.357 \\
50 → 14 & & & & 50 → 15 & & & \\
natCu(a,x)\textsuperscript{66}Ga & 0.00063584 & 0.00061618 & -3.091 & natCu(a,x)\textsuperscript{66}Ga & 0.00030845 & 0.00029724 & -3.635 \\
71.1 → 7.4 & & & & 47.5 → 8 & & & \\
natCu(a,x)\textsuperscript{67}Ga & 0.00113341 & 0.00110339 & -2.648 & natCu(a,x)\textsuperscript{67}Ga & 0.00030793 & 0.00029759 & -3.357 \\
100 → 14.4 & & & & 50 → 15 & & & \\
natTi(a,x)\textsuperscript{51}Cr & 0.00062416 & 0.00062114 & -0.484 & natTi(a,x)\textsuperscript{51}Cr & 0.00062416 & 0.00062114 & -0.484 \\
50 → 5 & & & & 50 → 5 & & & \\
natAl(\textsuperscript{3}He,x)\textsuperscript{22}Na & 0.00619959 & 0.00620580 & 0.100 & natAl(\textsuperscript{3}He,x)\textsuperscript{22}Na & 0.00386959 & 0.00386478 & -0.124 \\
130 → 8 & & & & 100 → 8 & & & \\
natAl(\textsuperscript{3}He,x)\textsuperscript{24}Na & 0.00690837 & 0.00691254 & 0.060 & natAl(\textsuperscript{3}He,x)\textsuperscript{24}Na & 0.00367837 & 0.00367900 & 0.017 \\
140 → 20 & & & & 99.9 → 20 & & & \\
natCu(\textsuperscript{3}He,x)\textsuperscript{63}Zn & 0.00032321 & 0.00031308 & -3.134 & - & - & - & - \\
44 → 9.3 & & & & & & & \\
natCu(\textsuperscript{3}He,x)\textsuperscript{65}Zn & 0.00130803 & 0.00127831 & -2.271 & - & - & - & - \\
95 → 8 & & & & & & & \\
natCu(\textsuperscript{3}He,x)\textsuperscript{66}Ga & 0.00042097 & 0.00040806 & -3.067 & - & - & - & - \\
50 → 6.4 & & & & & & & \\
natTi(\textsuperscript{3}He,x)\textsuperscript{48}V & 0.00479252 & 0.00482475 & 0.673 & natTi(\textsuperscript{3}He,x)\textsuperscript{48}V & 0.00263508 & 0.00265200 & 0.642 \\
140 → 5 & & & & 100 → 4.5 & & & \\
natAl(d,x)\textsuperscript{22}Na & 0.02057813 & 0.02055290 & -0.123 & natAl(d,x)\textsuperscript{22}Na & 0.01467000 & 0.01465380 & -0.110 \\
99.9 → 17 & & & & 83.9 → 18.5 & & & \\
natAl(d,x)\textsuperscript{24}Na & 0.02117093 & 0.02114440 & -0.125 & natAl(d,x)\textsuperscript{24}Na & 0.01705060 & 0.01701870 & -0.187 \\
100 → 10 & & & & 88.4 → 7.8 & & & \\
natCu(d,x)\textsuperscript{62}Zn & 0.00285701 & 0.00279363 & -2.218 & - & - & - & - \\
60 → 17 & & & & & & & \\
natCu(d,x)\textsuperscript{63}Zn & 0.00315173 & 0.00307295 & -2.500 & - & - & - & - \\
60 → 6.5 & & & & & & & \\
natCu(d,x)\textsuperscript{65}Zn & 0.00272863 & 0.00265260 & -2.786 & - & - & - & - \\
54.9 → 4.3 & & & & & & & \\
natFe(d,x)\textsuperscript{56}Co & 0.00240778 & 0.00237764 & -1.252 & natFe(d,x)\textsuperscript{56}Co & 0.00239778 & 0.00236564 & -1.341 \\
50 → 8 & & & & 49.9 → 8 & & & \\
natNi(d,x)\textsuperscript{56}Co & 0.00262253 & 0.00255443 & -2.597 & - & - & - & - \\
55 → 2.5 & & & & & & & \\
natNi(d,x)\textsuperscript{58}Co & 0.00238253 & 0.00231321 & -2.909 & - & - & - & - \\
52 → 2.5 & & & & & & & \\
natNi(d,x)\textsuperscript{61}Cu & 0.00238725 & 0.00231756 & -2.921 & natNi(d,x)\textsuperscript{61}Cu & 0.00209496 & 0.00204139 & -2.557 \\
52 → 2 & & & & 48.4 → 2.3 & & & \\
natTi(d,x)\textsuperscript{46}Sc & 0.00974110 & 0.00977496 & 0.348 & - & - & - & - \\
80 → 3 & & & & & & & \\
natTi(d,x)\textsuperscript{48}V & 0.00621906 & 0.00622643 & 0.118 & natTi(d,x)\textsuperscript{48}V & 0.00411906 & 0.00411692 & -0.052 \\
62 → 2 & & & & 49.1 → 2 & & & \\
natAl(p,x)\textsuperscript{22}Na & 1.51694000 & 1.51259000 & -0.287 & natAl(p,x)\textsuperscript{22}Na & 0.03364000 & 0.03371680 & 0.228 \\
1000 → 24.7 & & & & 100 → 25.2 & & & \\
natAl(p,x)\textsuperscript{24}Na & 0.03388000 & 0.03397950 & 0.294 & natAl(p,x)\textsuperscript{24}Na & 0.03368000 & 0.03376130 & 0.241 \\
100 → 24 & & & & 100 → 25 & & & \\
natCu(p,x)\textsuperscript{56}Co & 0.01019000 & 0.01013320 & -0.558 & natCu(p,x)\textsuperscript{56}Co & 0.01056000 & 0.01048760 & -0.686 \\
100 → 43 & & & & 100 → 40 & & & \\
natCu(p,x)\textsuperscript{58}Co & 0.01214000 & 0.01205440 & -0.705 & - & - & - & - \\
100 → 23.7 & & & & & & & \\
natCu(p,x)\textsuperscript{62}Zn & 0.01283002 & 0.01272550 & -0.815 & natCu(p,x)\textsuperscript{62}Zn & 0.00494477 & 0.00488849 & -1.138 \\
100 → 13 & & & & 59.6 → 13.5 & & & \\
natCu(p,x)\textsuperscript{63}Zn & 0.01315602 & 0.01303820 & -0.896 & natCu(p,x)\textsuperscript{63}Zn & 0.00386942 & 0.00381483 & -1.411 \\
100 → 4 & & & & 50 → 4.3 & & & \\
natCu(p,x)\textsuperscript{65}Zn & 0.01318879 & 0.01306870 & -0.911 & natCu(p,x)\textsuperscript{65}Zn & 0.01317879 & 0.01304590 & -1.008 \\
100 → 2.2 & & & & 99.9 → 2.2 & & & \\
natMo(p,x)\textsuperscript{96}Tc & 0.00371416 & 0.00371741 & 0.088 & - & - & - & - \\
50 → 4.5 & & & & & & & \\
natNi(p,x)\textsuperscript{57}Ni & 0.01228554 & 0.01218250 & -0.839 & natNi(p,x)\textsuperscript{57}Ni & 0.00337958 & 0.00334454 & -1.037 \\
100 → 12 & & & & 50 → 13 & & & \\
natTi(p,x)\textsuperscript{46}Sc & 0.01632326 & 0.01650950 & 1.141 & - & - & - & - \\
80 → 7 & & & & & & & \\
natTi(p,x)\textsuperscript{48}V & 0.02440440 & 0.02469800 & 1.203 & natTi(p,x)\textsuperscript{48}V & 0.00709533 & 0.00716132 & 0.930 \\
100 → 3.8 & & & & 50 → 5 & & & \\
\bottomrule
\end{tabular}
\end{table}

\begin{table}[htbp]
\centering
\scriptsize
\setstretch{1}
\caption{Percentage Deviation of Calculated Activities from IAEA BMR 2007 and 2017 Benchmarks}
\label{tab:iaea_bmr_comparison}
\begin{tabular}{p{2.3cm}ccc p{2.3cm}ccc}
\toprule
\multicolumn{4}{c}{\textbf{IAEA BMR 2017 DATABASE}} & \multicolumn{4}{c}{\textbf{IAEA BMR 2007 DATABASE}} \\
\cmidrule(r){1-4} \cmidrule(l){5-8}
\multirow{2}{*}{\textbf{Reaction}} & \multicolumn{2}{c}{\textbf{Product Activity (MBq)}} & \multirow{2}{*}{\textbf{\% Dif.}} & \multirow{2}{*}{\textbf{Reaction}} & \multicolumn{2}{c}{\textbf{Product Activity (Mbq)}} & \multirow{2}{*}{\textbf{\% Dif.}} \\
\cmidrule(r){2-3} \cmidrule(l){6-7}
 & \textbf{IAEA} & \textbf{This Work} & & & \textbf{IAEA} & \textbf{This Work} & \\
\midrule
natAl(a,x)\textsuperscript{22}Na & 0.270124 & 0.135378 & -49.88 & natAl(a,x)\textsuperscript{22}Na & 0.050572 & 0.050204 & -0.73 \\
150 → 31 & & & & 100 → 30 & & & \\
natAl(a,x)\textsuperscript{24}Na & 404.521447 & 202.722 & -49.90 & natAl(a,x)\textsuperscript{24}Na & 67.313206 & 65.541 & -2.63 \\
169.9 → 31 & & & & 100 → 30 & & & \\
natCu(a,x)\textsuperscript{65}Zn & 0.744198 & 0.357833 & -51.92 & natCu(a,x)\textsuperscript{65}Zn & 0.405087 & 0.388138 & -4.18 \\
50 → 14 & & & & 50 → 15 & & & \\
natCu(a,x)\textsuperscript{66}Ga & 235.284081 & 115.343 & -50.98 & natCu(a,x)\textsuperscript{66}Ga & 84.433179 & 81.8575 & -3.05 \\
71.1 → 7.4 & & & & 47.5 → 8 & & & \\
natCu(a,x)\textsuperscript{67}Ga & 20.532650 & 9.87541 & -51.90 & natCu(a,x)\textsuperscript{67}Ga & 9.841254 & 9.40851 & -4.40 \\
100 → 14.4 & & & & 50 → 15 & & & \\
natTi(a,x)\textsuperscript{51}Cr & 2.968656 & 1.46876 & -50.53 & natTi(a,x)\textsuperscript{51}Cr & 1.466325 & 1.44897 & -1.18 \\
50 → 5 & & & & 50 → 5 & & & \\
natAl(\textsuperscript{3}He,x)\textsuperscript{22}Na & 0.292998 & 0.146741 & -49.92 & natAl(\textsuperscript{3}He,x)\textsuperscript{22}Na & 0.081951 & 0.081126 & -1.01 \\
130 → 8 & & & & 100 → 8 & & & \\
natAl(\textsuperscript{3}He,x)\textsuperscript{24}Na & 270.513352 & 135.787 & -49.80 & natAl(\textsuperscript{3}He,x)\textsuperscript{24}Na & 79.068735 & 76.728 & -2.96 \\
140 → 20 & & & & 99.9 → 20 & & & \\
natCu(\textsuperscript{3}He,x)\textsuperscript{63}Zn & 1827.441107 & 886.383 & -51.50 & - & - & - & - \\
44 → 9.3 & & & & & & & \\
natCu(\textsuperscript{3}He,x)\textsuperscript{65}Zn & 0.803565 & 0.390913 & -51.35 & - & - & - & - \\
95 → 8 & & & & & & & \\
natCu(\textsuperscript{3}He,x)\textsuperscript{66}Ga & 34.954440 & 17.1272 & -51.00 & - & - & - & - \\
50 → 6.4 & & & & & & & \\
natTi(\textsuperscript{3}He,x)\textsuperscript{48}V & 29.122568 & 14.6174 & -49.80 & natTi(\textsuperscript{3}He,x)\textsuperscript{48}V & 11.386933 & 11.299 & -0.77 \\
140 → 5 & & & & 100 → 4.5 & & & \\
natAl(d,x)\textsuperscript{22}Na & 0.703787 & 0.672188 & -4.49 & natAl(d,x)\textsuperscript{22}Na & 0.441608 & 0.44039 & -0.28 \\
99.9 → 17 & & & & 83.9 → 18.5 & & & \\
natAl(d,x)\textsuperscript{24}Na & 953.139797 & 911.251 & -4.39 & natAl(d,x)\textsuperscript{24}Na & 748.816933 & 745.74 & -0.41 \\
100 → 10 & & & & 88.4 → 7.8 & & & \\
natCu(d,x)\textsuperscript{62}Zn & 223.991610 & 219.37 & -2.06 & - & - & - & - \\
60 → 17 & & & & & & & \\
natCu(d,x)\textsuperscript{63}Zn & 10439.349766 & 10159.4 & -2.68 & - & - & - & - \\
60 → 6.5 & & & & & & & \\
natCu(d,x)\textsuperscript{65}Zn & 1.287463 & 1.24678 & -3.16 & - & - & - & - \\
54.9 → 4.3 & & & & & & & \\
natFe(d,x)\textsuperscript{56}Co & 5.664871 & 5.35127 & -5.54 & natFe(d,x)\textsuperscript{56}Co & 5.310495 & 5.25937 & -0.96 \\
50 → 8 & & & & 49.9 → 8 & & & \\
natNi(d,x)\textsuperscript{56}Co & 3.775469 & 3.69907 & -2.02 & - & - & - & - \\
55 → 2.5 & & & & & & & \\
natNi(d,x)\textsuperscript{58}Co & 7.155222 & 6.95622 & -2.78 & - & - & - & - \\
52 → 2.5 & & & & & & & \\
natNi(d,x)\textsuperscript{61}Cu & 402.867636 & 388.85 & -3.48 & natNi(d,x)\textsuperscript{61}Cu & 381.934402 & 370.715 & -2.94 \\
52 → 2 & & & & 48.4 → 2.3 & & & \\
natTi(d,x)\textsuperscript{46}Sc & 7.960974 & 7.61876 & -4.30 & - & - & - & - \\
80 → 3 & & & & & & & \\
natTi(d,x)\textsuperscript{48}V & 42.243232 & 40.2914 & -4.62 & natTi(d,x)\textsuperscript{48}V & 33.220512 & 33.324 & 0.31 \\
62 → 2 & & & & 49.1 → 2 & & & \\
natAl(p,x)\textsuperscript{22}Na & 27.107384 & 24.4037 & -9.97 & natAl(p,x)\textsuperscript{22}Na & 0.959248 & 0.919922 & -4.10 \\
1000 → 24.7 & & & & 100 → 25.2 & & & \\
natAl(p,x)\textsuperscript{24}Na & 546.281143 & 508.331 & -6.95 & natAl(p,x)\textsuperscript{24}Na & 543.810435 & 505.863 & -6.98 \\
100 → 24 & & & & 100 → 25 & & & \\
natCu(p,x)\textsuperscript{56}Co & 1.898706 & 1.77977 & -6.27 & natCu(p,x)\textsuperscript{56}Co & 2.064601 & 1.77724 & -13.93 \\
100 → 43 & & & & 100 → 40 & & & \\
natCu(p,x)\textsuperscript{58}Co & 11.800518 & 11.13 & -5.68 & - & - & - & - \\
100 → 23.7 & & & & & & & \\
natCu(p,x)\textsuperscript{62}Zn & 609.590150 & 593.735 & -2.60 & natCu(p,x)\textsuperscript{62}Zn & 458.064603 & 435.492 & -4.93 \\
100 → 13 & & & & 59.6 → 13.5 & & & \\
natCu(p,x)\textsuperscript{63}Zn & 12342.907007 & 11891 & -3.66 & natCu(p,x)\textsuperscript{63}Zn & 9080.839240 & 8823.91 & -2.83 \\
100 → 4 & & & & 50 → 4.3 & & & \\
natCu(p,x)\textsuperscript{65}Zn & 0.850647 & 0.823464 & -3.20 & natCu(p,x)\textsuperscript{65}Zn & 0.856699 & 0.829151 & -3.22 \\
100 → 2.2 & & & & 99.9 → 2.2 & & & \\
natMo(p,x)\textsuperscript{96}Tc & 105.078320 & 100.619 & -4.24 & - & - & - & - \\
50 → 4.5 & & & & & & & \\
natNi(p,x)\textsuperscript{57}Ni & 1026.755781 & 983.9 & -4.17 & natNi(p,x)\textsuperscript{57}Ni & 380.473131 & 375.232 & -1.38 \\
100 → 12 & & & & 50 → 13 & & & \\
natTi(p,x)\textsuperscript{46}Sc & 8.672426 & 8.31169 & -4.16 & - & - & - & - \\
80 → 7 & & & & & & & \\
natTi(p,x)\textsuperscript{48}V & 45.221928 & 43.212 & -4.44 & natTi(p,x)\textsuperscript{48}V & 30.994762 & 31.1354 & 0.45 \\
100 → 3.8 & & & & 50 → 5 & & & \\
\bottomrule
\end{tabular}
\end{table}

\begin{table}[htbp]
\centering
\caption{Statistical comparison of IMRA model results with IAEA Beam Monitor Reaction (BMR) databases.}
\begin{tabular}{lcccc}
\hline
\textbf{} & \multicolumn{2}{c}{\textbf{IAEA BMR 2017 Database}} & \multicolumn{2}{c}{\textbf{IAEA BMR 2007 Database}} \\ 
\cline{2-5}
\textbf{Metric} & \textbf{Range} & \textbf{Activity} & \textbf{Range} & \textbf{Activity}\\ 
\hline
PEARSON & 0.9999999346 & 0.9982495545 & 0.9999894297 & 0.9999934307 \\ 
MAPE & 1.0657 & 4.3125$^\ast$ & 1.0282 & 2.8916 \\ 
STDEV.P (\%Diff.) & 1.3906 & 1.7619$^\ast$ & 1.1357 & 2.9794 \\ 
\hline
\multicolumn{5}{l}{$^\ast$ Analysis based on 22 reactions after excluding 12 cases with factor-of-two deviations.} \\
\end{tabular}
\label{tab:stats_iaea_imra}
\end{table}

\section{Conclusion}
This study originated from the reliability testing of an original computational tool (IMRA) developed within the framework of a master’s thesis. During this validation process, activity values calculated for a set of alpha- and helium-3-induced monitor reactions were compared with reference data reported in the IAEA--BMR databases.

For twelve monitor reactions, the activity values reported in the IAEA--BMR 2017 dataset were found to differ systematically from the corresponding values obtained using the present computational approach, with deviations on the order of approximately a factor of two. In contrast, for the same reactions, the activity values reported in the IAEA--BMR 2007 data set showed good agreement with the present calculations.

A detailed examination indicates that these differences may be related to the treatment of beam flux normalization in the activity calculation procedure. For charged-particle-induced reactions, the particle flux $\phi$ is commonly expressed as
\begin{equation}
\phi = \frac{I}{e \, z},
\end{equation}
where $\phi$ is the particle flux, $I$ is the beam current, $e$ is the elementary charge ($1.602 \times 10^{-19}$~C), and $z$ is the charge number of the projectile. Variations in the implementation of this normalization can directly propagate into calculated activity values.

The absence of comparable deviations in the IAEA--BMR 2007 data set for the same reactions suggests that the observed differences are not intrinsic to the activation formalism itself, but may be associated with dataset-specific evaluation or processing choices. These observations are not related to the evaluated cross-section data themselves, which remain consistent within the IAEA--BMR framework.

The observations reported in this study are presented as independent computational findings obtained during the validation of an original calculation tool. The results presented here are provided as part of the authors' scientific responsibility to contribute independent computational verification results to the nuclear data community. Independent validation studies of this type can serve as complementary inputs to ongoing database evaluation and update efforts carried out by the responsible international expert groups.

\subsection*{Notes on Minor Data Inconsistencies}

The following observations are reported solely for completeness and transparency
and do not affect the main results or conclusions of the present study. These
dataset-level inconsistencies were identified during the independent verification
process and are documented here as part of the authors’ scientific responsibility.

\begin{itemize}
    \item In two reactions (IAEA BMR 2017 dataset: natAl($^3$He,x)$^{22}$Na and natTi(p,x)$^{46}$Sc), a discrepancy was identified due to the misuse of activity units. Although the dataset specifies irradiation in microampere units, the corresponding activity values appear to have been calculated assuming milliampere irradiation.
    
    \item In three reactions (IAEA BMR 2007 dataset: natCu(p,x)$^{62}$Zn, natCu(p,x)$^{65}$Zn, natNi(p,x)$^{57}$Ni), the physical yield activity column and the activity values reported after 1~hour of irradiation were inadvertently interchanged.
\end{itemize}

These minor inconsistencies are not central to the scope of this work and do not
influence the comparative assessment of the activity calculation methodology
presented in this study.

\section*{Acknowledgements}
This work extends the research presented in the master's thesis \textit{``Analysis of charged particle activation used in positron emitter radioisotope production with unique computer simulations''} (İTÜ, 2020), supervised by Prof. Dr. İskender Atilla Reyhancan. The authors acknowledge the foundational work conducted during this thesis.

AI-based tools were used solely for language editing and LaTeX formatting. The corresponding author retains full responsibility for the scientific content of the manuscript.
\newpage
\bibliographystyle{IEEEtran}
\bibliography{BMR}

\end{document}